\documentstyle[12pt,aaspp4]{article}
\begin{document}
\newcommand{\up}[1]{\ifmmode^{\rm #1}\else$^{\rm #1}$\fi}
\newcommand{\zdot}{\makebox[0pt][l]{.}}
\newcommand{\upd}{\up{d}}
\newcommand{\uph}{\up{h}}
\newcommand{\upm}{\up{m}}
\newcommand{\ups}{\up{s}}
\newcommand{\arcd}{\ifmmode^{\circ}\else$^{\circ}$\fi}
\newcommand{\arcm}{\ifmmode{'}\else$'$\fi}
\newcommand{\arcs}{\ifmmode{''}\else$''$\fi}

\title{The Araucaria Project. An Accurate Distance to the Local Group Galaxy NGC 6822 
from Near-Infrared Photometry of Cepheid Variables
\footnote{Based on observations obtained with the ESO NTT telescope for 
Large Programme 171.D-0004, and with the Magellan Telescope of Las Campanas Observatory
}
}
\author{Wolfgang Gieren}
\affil{Universidad de Concepci{\'o}n, Departamento de Fisica, Astronomy Group,
Casilla 160-C, Concepci{\'o}n, Chile}
\authoremail{wgieren@astro-udec.cl}
\author{Grzegorz Pietrzy{\'n}ski}
\affil{Universidad de Concepci{\'o}n, Departamento de Fisica, Astronomy
Group,
Casilla 160-C,
Concepci{\'o}n, Chile}
\affil{Warsaw University Observatory, Al. Ujazdowskie 4, 00-478, Warsaw,
Poland}
\authoremail{pietrzyn@hubble.cfm.udec.cl}
\author{Krzysztof Nalewajko}
\affil{Warsaw University Observatory, Al. Ujazdowskie 4, 00-478, Warsaw,
Poland}
\authoremail{knalewaj@astrouw.edu.pl}
\author{Igor Soszy{\'n}ski}
\affil{Universidad de Concepci{\'o}n, Departamento de Fisica, Astronomy Group, 
Casilla 160-C, Concepci{\'o}n, Chile}
\affil{Warsaw University Observatory, Al. Ujazdowskie 4, 00-478, Warsaw,
Poland}
\authoremail{soszynsk@astro-udec.cl}
\author{Fabio Bresolin}
\affil{Institute for Astronomy, University of Hawaii at Manoa, 2680 Woodlawn 
Drive, 
Honolulu HI 96822, USA}
\authoremail{bresolin@ifa.hawaii.edu}
\author{Rolf-Peter Kudritzki}
\affil{Institute for Astronomy, University of Hawaii at Manoa, 2680 Woodlawn 
Drive,
Honolulu HI 96822, USA}
\authoremail{kud@ifa.hawaii.edu}
\author{Dante Minniti}
\affil{Pontificia Universidad Cat{\'o}lica de Chile, Departamento de Astronomia y Astrofisica,
Casilla 306, Santiago 22, Chile}
\authoremail{dante@astro.puc.cl}
\author{Aaron Romanowsky}
\affil{Universidad de Concepci{\'o}n, Departamento de Fisica, Astronomy
Group, Casilla 160-C, Concepci{\'o}n, Chile}
\authoremail{romanow@astro-udec.cl}

\begin{abstract}
We have measured near-infrared magnitudes in the J and K bands for
56 Cepheid variables in the Local Group galaxy NGC 6822 with well-determined
periods and optical light curves in the V and I bands. Using the template light curve approach of
Soszy{\'n}ski, Gieren and Pietrzy{\'n}ski, accurate mean magnitudes were obtained
from these data which allowed us to determine with unprecedented accuracy the distance to NGC 6822 
from a multi-wavelength period-luminosity solution in the VIJK bands.
From our data,
we obtain a distance to NGC 6822 of $(m-M)_{0} = 23.312 \pm 0.021$ (random error) mag,
with an additional systematic uncertainty of $\sim$3\%. This distance value is tied to
an assumed LMC distance modulus of 18.50.
From our multiwavelength
approach, we find for the total (average) reddening to the NGC 6822 Cepheids 
$E(B-V) = 0.356 \pm 0.013$ mag, which is in excellent agreement with a previous determination
of McGonegal et al. from near-infrared photometry
and implies significant internal reddening of the Cepheids in NGC 6822. Our
present, definitive distance determination of NGC 6822 from Cepheids agrees within 2\% with
the previous distance we had derived from optical photometry alone,
but has significantly reduced error bars. 
 
Our Cepheid distance to NGC 6822 is in excellent agreement with the recent independent determination
of Cioni \& Habing from the I-band magnitude of the tip of the red giant branch. It also agrees well,
within the errors,
with the early determination of McGonegal et al. (1983) from random-phase H-band photometry
of nine Cepheids.
\end{abstract}


\section{Introduction}
Cepheid variables are the most important distance indicators for calibrating the first
rungs of the distance ladder, out to some 30 Mpc. As young stars, Cepheids tend to lie 
in dusty regions in their spiral or irregular parent galaxies. This implies that
distance determinations using Cepheids in the optical spectral range are very
sensitive to an accurate knowledge of the total reddening, including the reddening
produced inside the parent galaxy itself. Consequently, efforts to use Cepheids
as distance indicators in near-infrared passbands have started as early as in
the eighties of the past century with the pioneering work of McGonegal et al. (1982).
Perhaps the most important obstacle to derive truly accurate distances to nearby
galaxies from infrared photometry of their Cepheids was the lack of well-calibrated
fiducial period-luminosity (PL) relations in the near-infrared JHK bands. This long-standing
problem was finally solved with the work of Persson et al. (2004) who provided
such fiducial PL relations in the LMC. Very recently, Gieren et al. (2005a) have also
provided well-calibrated Cepheid PL relations in JHK for the Milky Way, which are
in very good agreement with the corresponding LMC relations when using an improved version
of their original infrared surface brightness technique (Fouqu{\'e} \& Gieren 1997;
Gieren, Fouqu{\'e} \& G{\'o}mez 1997, 1998) employed for the distance determination of
individual Cepheids.

In the {\it Araucaria Project}, we have conducted optical (BVI) surveys for Cepheid variables
from wide-field images for a number of nearby galaxies in the Local Group, and in the
more distant Sculptor Group. The main goal is to carry out a detailed investigation of
the respective dependences on environmental properties of a number of stellar distance
indicators, including Cepheid variables, hereby improving the use of these objects 
as standard candles. While we discover the Cepheids in optical bands where these stars
are easy to find due to their relatively large light amplitudes and characteristic light curve
shapes, the most accurate distance determinations come from near-infrared
follow-up photometry of selected subsamples of Cepheids, which allow a dramatic reduction
of the systematic uncertainty of the distances due to reddening. Our previous near-infrared work
on Cepheids in the Araucaria Project in NGC 300 (Gieren et al. 2005b), IC 1613 (Pietrzy{\'n}ski
et al. 2006) and NGC 3109 (Soszy{\'n}ski et al. 2006) has already clearly demonstrated that
errors in the adopted reddenings, which often neglect the contribution to the
total reddening produced inside the host galaxies, is usually the largest single
systematic error in Cepheid distances to galaxies derived from optical photometry alone.
Near-infrared follow-up work is a {\it must} if one wants to push
the distance uncertainties down to the $\sim$3\% level.

In the present paper, we are presenting near-infrared photometry in the J and K bands
for 56 Cepheid variables with periods between 124 and 1.7 days which are scattered across
the entire surface of the Local Group irregular galaxy NGC 6822. These variables were previously
discovered from optical VI images by Pietrzy{\'n}ski et al. (2004; hereafter Paper I). 
At a galactic latitude of -18 degrees, there is substantial foreground reddening
to NGC 6822 which implies that not only any intrinsic contribution to the reddening
of the Cepheids must be carefully determined for an accurate distance measurement, 
but that the foreground reddening may not be well determined either.
We will show that
as in our previous work on NGC 300, IC 1613 and NGC 3109, the combination of the new
infrared photometry presented here with the previous optical photometry of the Cepheids
results in a very accurate determination of the total reddening, and of the distance
to NGC 6822, which will be extremely useful for the study of the environmental dependences
of Cepheids and other stellar distance indicators by comparative studies, which will be
the subject of forthcoming papers in our project. Our paper is organized as follows: in
section 2, we describe the observations, reduction and calibrations of our data; in
section 3, we present the calibrated near-infrared mean magnitudes of the Cepheids in
our selected fields and determine the total reddening, and distance of NGC 6822; and in
section 4, we discuss our results and present some conclusions.

\section{Observations, Data Reduction and Calibration}
The near-infrared data presented in this paper were collected with two instruments:
the PANIC camera at the Magellan-Baade telescope at Las Campanas Observatory, and
the SOFI camera at the ESO NTT telescope on La Silla. The field of view of the
PANIC near-infrared camera is about 2 x 2 arcmin, and the pixel scale is 0.15 arcsec/pixel.
The SOFI infrared camera was used in its Large Field setup, with a field of view
of 4.9 x 4.9 arcmin, and a scale of 0.288 arcsec/pixel. More details on these instruments
can be found on the respective webpages. We observed a total of 8 different PANIC fields,
and 6 SOFI fields. The fields are overlapping and cover most of the spatial extent
of NGC 6822. The location of the different fields was chosen as to maximize the number
of the (relatively few) long-period Cepheids in the fields; with our chosen field coordinates,
we were able to obtain infrared photometry for 17 out of the 22 Cepheids with periods
longer than 10 days in the Cepheid catalog we presented in Paper I. Fig. 1
shows the location of our observed fields in NGC 6822.

Single deep J and Ks observations were obtained under good seeing conditions on
4 different  nights at Las Campanas, and 2 nights on La Silla. On two
photometric nights, we observed a large number of photometric standard stars 
from the UKIRT system (Hawarden et al. 2001). In order to account for the rapid sky brightness 
variations in the near-infrared, the observations were performed with a jittering technique.
In the Ks filter, we obtained six consecutive 10 s integrations (DITs) at a given
sky position and then moved the telescope by about 20 arcsec to a different random position.
Integrations at 62 different jittering positions resulted in a total net exposure 
time of 62 minutes in this filter, for a given field. In the case of the J filter,
in which the sky level variations are less pronounced than in K, two consecutive
20 s exposures were obtained at each of 25 jittering positions, which resulted in
a total net exposure of about 17 minutes, for any given field.

Sky subtraction was performed by using a two-step process employing the masking of stars
with the XDIMSUM IRAF\footnote{IRAF is distributed by the
National Optical Astronomy Observatories, which are operated by the
Association of Universities for Research in Astronomy, Inc., under
cooperative agreement with the NSF.} package in an analogous manner as described in Pietrzy{\'n}ski 
and Gieren (2002). Then the single images were flatfielded and stacked into the final
images. PSF photometry was obtained using DAOPHOT and ALLSTAR, following the procedure
described in Pietrzy{\'n}ski, Gieren and Udalski (2002). In order to derive the
aperture corrections for each frame, about 7-10 relatively isolated and bright stars 
were selected, and all neighboring stars were removed using an iterative procedure.
Finally, we measured the aperture magnitudes for the selected stars with the DAOPHOT
program using apertures of 16 pixels. The median from the differences between the
aperture magnitudes obtained this way, and the corresponding PSF magnitudes, averaged
over all selected stars was finally adopted as the aperture correction for a given frame.
The rms scatter from all measurements was always smaller than 0.02 mag.

In order to accurately transform our SOFI data to the standard system, a large number (between
8 and 15) of standard stars from the UKIRT system (Hawarden et al. 2001) was observed
under photometric conditions at a variety of airmasses, together with our science fields.
The standard stars were chosen to have colors bracketting the colors of the Cepheids
in NGC 6822. The aperture photometry for our standard stars was performed with DAOPHOT
using the same aperture as for the calculation of the aperture corrections. Given the
relatively large number of standard stars we observed on each night, the transformation
coefficients were derived on every night. The accuracy of the zero points of our photometry
was determined to be $\pm$ 0.02 mag. The data obtained with the PANIC camera
were transformed to the standard system using stars in common to the
SOFI fields.

Since our science fields overlap (see Fig. 1), we were able to perform an internal check on
our photometry by comparing the magnitudes of stars located in the common regions. In 
all cases, the independently calibrated magnitudes agree within 0.02 - 0.03 mag in both,
J and K filters. For the relatively bright stars in our fields, the magnitudes can be
compared to the corresponding 2MASS magnitudes. Fig. 2 presents such a comparison in
J and K for the common bright stars. Before calculating the differences between our own,
and the 2MASS photometry, we transformed our photometry, which had been calibrated onto the
UKIRT system, to the 2MASS system using the equations provided by Carpenter (2001). 
Fig. 2 suggests that our magnitudes in both J and K are about 0.03 mag brighter than
the corresponding 2MASS magnitudes, but given the rather large scatter in the 2MASS data
for the fainter stars this offset does not seem significant. In any case, the comparison
confirms our conclusion that our photometric zero points in J and K are accurate
to at least 0.03 mag.

The pixel positions of the stars were transformed to the equatorial coordinate system
using the Digital Sky Survey (DSS) images. For this purpose, we used the algorithm 
developed and used in the OGLE Project (Udalski et al. 1998). The accuracy of our
astrometric transformations is better than 0.3 arcsec.

\section{Results}
\subsection{The Cepheid mean J and K magnitudes}
Alltogether, our observed fields in NGC 6822 contain 56 Cepheids bright enough to
measure their near-infrared magnitudes. As we mentioned
before, we chose our fields in such a way as to maximize the number of long-period
Cepheids in them, and we obtained photometry for 17 of the 22 Cepheids with periods
longer than 10 days in our previous catalog (Paper I). The
faintest Cepheids for which we could measure relatively accurate J and K magnitudes 
have periods around 4-5 days. 

In Table 1, we present the journal of individual observations of these 56 Cepheid variables.
The identifications are as in the catalog in Paper I.
Nearly all of these objects have observations at more than one pulsation phase, and some of
them have JK observations at four different phases in their light curves. The mean magnitudes
of the Cepheids were derived from each individual single-phase measurement from the
template light curve method developed by our group (Soszy{\'n}ski et al. 2005) which
uses the V- and I-band phases of the individual near-IR observations, and the light curve amplitudes
in V and I to calculate the differences of the individual single-phase magnitudes to
the mean magnitudes in J and K. For a detailed description of the technique, the reader
is referred to this paper in which we demonstrated that the mean K and J magnitudes of Cepheids
can be derived from just one single-phase infrared observation with an accuracy of 0.02-0.03 mag 
provided that high-quality optical light curves and periods are available for the variables, 
which is clearly the case in this study. Obviously, the availability of more than one
near-IR observation of a Cepheid increases the accuracy of the determination of its mean
near-IR magnitude because one can average the various independent determinations.
In Table 2, we present the final intensity-mean JK magnitudes of our
Cepheid sample
with their estimated uncertainties, which were determined from the photometric
accuracies of the individual random-phase measurements of a given variable, and from the 
additional uncertainty introduced
in the calculation of the mean magnitudes from the template method. In all cases where
more than one near-IR observation for a given Cepheid was available, the independent
determinations of its mean magnitude agreed very well, typically within 1-3\%, reconfirming
the quality of our present near-infrared photometry as well as the accuracy of our
previous VI light curves and periods given in Paper I.

\subsection{Near-infrared period-luminosity relations and distance determination}
In Fig. 3, we show the J- and K-band Cepheid PL relations in NGC 6822 obtained from the mean magnitudes
and periods in Table 2. Our sample is seen to define the PL relations exceedingly well.
We have adopted a period cutoff of 4.8 days in order to avoid any significant contamination
of our sample with possible overtone pulsators which may appear in increasing numbers at shorter periods,
and which could bias the distance determination. Another reason to exclude stars with periods less
than 4.8 days is the increasingly larger random errors in their photometry due
to the increasing faintness of these objects.
For the fits to a line, we decided to exclude four objects which
are identified by open circles in Fig. 3. As in our discussion in our previous optical
work in Paper I, we exclude the one Cepheid with a period longer
than 100 days, for the reasons given in that paper. The other 3 objects (variables cep025,
028 and 052) are excluded on the basis of their large deviations from the mean J- and K-band
PL relations defined by the remaining 33 stars. The same 3 Cepheids deviate
consistently in both bands, in the sense that their magnitudes are too bright for
their periods, which suggests that these objects are blended with relatively bright
nearby stars which are not resolved in our images. One of these objects, cep028 at
logP=0.858, is also clearly overluminous in the optical (VI) PL relations whereas cep025
and cep052 lie close to the ridge lines in the V- and I-band PL relations. This could indicate
that for these two stars, the unresolved companions are very red and their contribution
to the measured fluxes of the Cepheids is only strongly noted in the near-infrared. For the
adopted sample of 33 Cepheids, the dispersions of the PL relations in both J and K
around the ridge lines are so small that they basically reflect the intrinsic dispersions
caused by the finite width of the Cepheid instability strip; this is suggested by a comparison
of the present relations to the exceedingly well-determined near-infrared Cepheid PL
relations in the LMC of Persson et al. (2004). Finally, we also note that a possible
contamination of our sample with first overtone pulsators is very unlikely, given that
the shortest periods of about 5 days in our sample are too long for first overtone Cepheids.

Least-squares fits to  a line to our data for the adopted 33 Cepheids in Fig. 3 yield
slopes of the PL relations of -3.28 $\pm$ 0.04 in K, and -3.23 $\pm$ 0.05 in J,
respectively. These values agree very well with the slopes for the Cepheid PL relations
in the LMC, which are -3.261 in K, and -3.153 in J (Persson et al. 2004), suggesting
that the slope of the PL relation in the near-infrared domain remains unchanged when
going from the LMC to the slightly more metal-poor NGC 6822 Cepheids. This result
is in line with the fact that the LMC slopes also yield very good fits to the Cepheids
in IC 1613 which are even more metal-poor than the Cepheids in NGC 6822 (Pietrzy{\'n}ski et al. 2006). 
Following the procedure we have adopted in our previous papers, we adopt the LMC
slopes of Persson et al. (2004) in our fits. This yields the following PL relations for NGC 6822: \\

J = -3.153 log P + (21.425 $\pm$ 0.025) \\

K = -3.261 log P + (20.999 $\pm$ 0.021) \\

These relations are shown in Fig. 3. It is seen that they fit the data extremely well.
Before calculating the relative distance of
NGC 6822 with respect to the LMC from the zero points in these relations,  
we need to convert our zero point magnitudes which are
calibrated onto the UKIRT system, to the NICMOS system on which the corresponding LMC
zero points were calibrated (Persson et al. 2004). According to Hawarden et al. (2001),
there are just zero point offsets between the UKIRT and NICMOS systems (e.g. no color dependences)
in the J and K filters, which amount to 0.034 and 0.015 mag, respectively. After adding
these offsets, and assuming an LMC true distance modulus of 18.50 as in our previous
work in the Araucaria Project, we derive distance moduli for NGC 6822 of 23.586 mag
in the J band, and 23.463 in the K band. Values for the distance modulus of 24.458 in V,
and 24.025 in I had been found in our previous optical work in Paper I.

As in our previous papers in this series (Gieren et al. 2005b; Pietrzy{\'n}ski et al. 2006),
we adopt the extinction law of Schlegel et al. (1998) and fit a straight line to the relation
$(m-M)_{0} = (m-M)_{\lambda} - A_{\lambda} = (m-M)_{\lambda} - E(B-V) * R_{\lambda}$.
The best least squares fit to this relation yields for the reddening, and the true
distance modulus of NGC 6822 the following values: \\

$(m-M)_{0} = 23.312 \pm 0.021$

$ E(B-V) = 0.356 \pm 0.013$

In Table 3 we give the adopted values of $R_{\lambda}$ and the unreddened distance moduli
in each band which are obtained with the reddening determined in our multi-wavelength
approach. The agreement between the unreddened distance moduli in each band is excellent.
In Fig. 4, we show the apparent distance moduli in VIJK as a function of $R_{\lambda}$, and the
best fit to the data; it is appreciated that the total reddening, and the true distance
modulus of NGC 6822 are indeed very well determined from this fit. Comparison of our reddening value
to the foreground reddening to NGC 6822 of 0.236 demonstrates that there is indeed a significant
contribution to the total reddening of the Cepheids which is produced by dust extinction in
the galaxy itself, in agreement with what we found for NGC 300 and IC 1613 in the earlier
papers in this series. In the case of NGC 6822, our data suggest that the reddening produced 
internal to the galaxy is 0.12 mag.

\section{Discussion and conclusions}
In the following, we will discuss the various assumptions we made, and possible systematic
errors which could affect our distance determination of NGC 6822.

The probably largest single systematic uncertainty on the derived distance modulus 
of NGC 6822 comes from the current uncertainty on the adopted distance modulus for the LMC.
This problem has been extensively discussed in the recent literature (e.g. Benedict et al. 2002),
and we will not add on this discussion here. We just note that our adopted value of 18.50
for the LMC distance is in agreement with the value adopted by the Key Project of 
Freedman et al. (2001), and since all the Cepheid distances obtained for the different target
galaxies in our project are tied to the same LMC distance, the {\it relative} distances
between our target galaxies are not affected.       

In a recent paper which presented our near-infrared Cepheid photometry for another Local Group
irregular galaxy, IC 1613 (Pietrzy{\'n}ski et al. 2006), we discussed in some detail why
our distance results derived from our current multi-wavelength approach are 
little affected by the choice of the fiducial PL relations in the LMC.
The Persson et al. (2004) JK Cepheid PL relations
we have used in this work are clearly extremely well established, at
least for periods longer than 10 days. The Persson et al. sample does
not contain, however, many Cepheids with shorter periods, so a possible
break in the PL relation at 10 days as advocated by Kanbur and Ngeow (2004), and
Ngeow et al. (2005), would not be easily detected. Our previous results in
the Araucaria Project on NGC 300 whose Cepheids have on average LMC
metallicity (Urbaneja et al. 2005), and on the much more metal-poor IC
1613 galaxy (Pietrzynski et al. 2006) have shown that within very small
uncertainties the PL relation in J and K seems universal in the
metallicity range from
about -0.3 to -1.0 dex. Any concerns about the universality of the PL
relation, as advocated by Sandage et al. (2004) and Ngeow and Kanbur
(2004), seem to refer to the metallicity range between solar and LMC abundance. While
he former studies find a steeper slope for the Milky Way PL relation
than for the LMC, in agreement with the work of Fouque et al. (2003) from the
infrared surface brightness technique, there is now recent evidence that
the Milky Way PL relation actually agrees in slope with the LMC relation
when a systematic error in the surface brightness technique is corrected
(Gieren et al. 2005a). Clearly, the question of the universality of the
PL relation, particularly in the range between solar and LMC metallicity,
has to be further investigated to obtain a clear-cut and convincing result, and
this is one of the main purposes of our project.
Regarding the possible break of the PL relation at 10 days discussed by
Sandage et al. (2004), and more recently by Ngeow et al. (2005), we
clearly have not seen such an effect in the data for any of the galaxies we have so far
studied in the Araucaria Project; the effect, if real, must be very subtle, and indeed
Ngeow et al. (2005) have estimated that its effect on distance determinations with the PL
relation should be less than 3\%, and therefore not a cause of serious concern as long
as we do not strive for distance accuracies of 1\%, which seem out of reach for
non-geometrical methods
at the present time.

As in the previous papers in this series on NGC 300 (Gieren et al. 2005b), IC 1613 (Pietrzy{\'n}ski
et al. 2006), and very recently NGC 3109 (Soszy{\'n}ski et al. 2006), we were able
to measure very accurate mean JK magnitudes for a very sizable sample of Cepheids in NGC 6822
which cover a broad range in periods, and establish a final sample of variables which
is very unlikely to be affected by the presence of overtone pulsators, or heavily blended stars which
could seriously bias our determination of the zero points of the PL relations in
the J and K bands. Regarding the possible effect of blending, viz. of the effect of
nearby companion stars to the Cepheids which are not resolved in our photometry, we have
shown in the case of the more distant NGC 300 from {\it Hubble Space Telescope} images
that the effect on the distance determination does not exceed 2\% (Bresolin et al. 2005).
In the present case of NGC 6822 the effect is expected to be smaller because the galaxy
is about 4 times closer than NGC 300, and has a smaller average stellar density. In
addition, heavily blended Cepheids can be rather easily recognized, and eliminated
from the sample used for the distance determination (as we did in this study) from their 
observed positions on
the PL planes {\it if} the observational scatter is low, as a consequence of using photometric data
of high quality. This is certainly the case for the present data, and we already remarked before
that the dispersion in Fig. 3 is so small that it approaches the {\it intrinsic} dispersion
expected from the finite width of the Cepheid instability strip, with only a small
additional contribution from photometric errors which is mostly visible at the
shortest periods in Fig. 3. This eliminates any concern that blending could bias our present
distance determination by more than 2\%. Also, the relatively large size of our sample 
eliminates concerns about the effect of incomplete filling of the instability strip (see also
the discussion in Pietrzy{\'n}ski et al. 2006, and our previous paper presenting the optical
photometry of the NGC 6822 Cepheids (Pietrzy{\'n}ski et al. 2004)). 

As we have shown in our previous papers in this series, the most important source of
uncertainty in Cepheid-based distance determinations of nearby, resolved galaxies
from optical photometric data alone is the interstellar reddening. In both NGC 300 and IC 1613
we found from our infrared studies that there is a very significant {\it intrinsic} contribution
to the reddening, in addition to the galactic foreground reddening in the directions to these galaxies.
The most important gain
in extending the Cepheid observations to the infrared is the possibility to determine
the total reddening very accurately, as again demonstrated in this paper, for the case
of NGC 6822. Our definitive distance determination of NGC 6822 from Cepheids presented
in this paper agrees very closely with our previous, preliminary result from VI photometry
presented in Paper I because we had happened to use the correct reddening value in that paper. 
The agreement of the reddening of NGC 6822 derived from Cepheid H-band photometry by 
McGonegal et al. (1983) with our present value from VIJK photometry of Cepheids
confirms the excellent pioneering infrared work done by these authors more than 20 years ago.

As a result of this discussion, and those presented in our previous papers in this series,
we conclude that the total effect of systematic errors due to the fiducial PL relations we use,
possible incomplete filling of the instability strip and contamination of our Cepheid
samples with overtone pulsators, and blending with unresolved companion stars in our
photometry does not exceed $\sim$3\%. The systematic uncertainty of our adopted photometric 
zero points, which propagates directly into the distance determination, is in the order of 0.03 mag,
or $\pm$1.5\%, in all bands. Random errors due to photometric noise are clearly less important
in the present case and do not contribute significantly to the error budget.
We therefore conclude that the present distance determination
to NGC 6822 from our combined optical and near-infrared photometry of Cepheids in this
galaxy is accurate, including systematics, to about $\pm$3 percent. We stress, however,
that this value does not include the contribution from the current uncertainty of the
LMC distance, which could be as large as 10\%. For the immediate purposes of the Araucaria
Project, however, {\it relative} distances are essential, and we have now another galaxy
whose distance relative to the LMC appears to be determined with the same high accuracy
of about $\pm$3\% as for the other galaxies for which we have used Cepheid infrared photometry
for distance determination, viz. NGC 300, IC 1613 and NGC 3109.

Finally, we note that our present distance result of 23.31 $\pm$ 0.02 (random) $\pm$$\sim$0.06 (systematic)
based on Cepheid variables agrees very well 
with the recent distance determination
of Cioni \& Habing (2005) who obtained a true distance modulus of 23.34 $\pm$ 0.12 mag for NGC 6822
from the observed I-band magnitude of the tip of the red giant branch. It also agrees reasonably
well with the former Cepheid near-infrared distance to NGC 6822 of $(m-M)_{0} = 23.47 \pm 0.11$
derived by McGonegal et al. (1983) using the same reddening as the one we found in this paper,
 from random-phase H-band photometry of 9 Cepheids which was not corrected to the mean magnitudes 
 as in our present work. This fact alone, combined with the small number of Cepheids they used
 in their PL solution may easily explain the 0.16 mag deviation of their value from ours.

\acknowledgments
WG, GP, DM and AR gratefully acknowledge financial support for this
work from the Chilean Center for Astrophysics FONDAP 15010003. 
Support from the DST and BW grants for 
Warsaw University Observatory is also acknowledged. It is a great pleasure
to thank the support astronomers at both ESO-La Silla and Las Campanas Observatories
for their expert help in the observations.

\begin{figure}[p] 
\vspace*{18cm}
\includegraphics{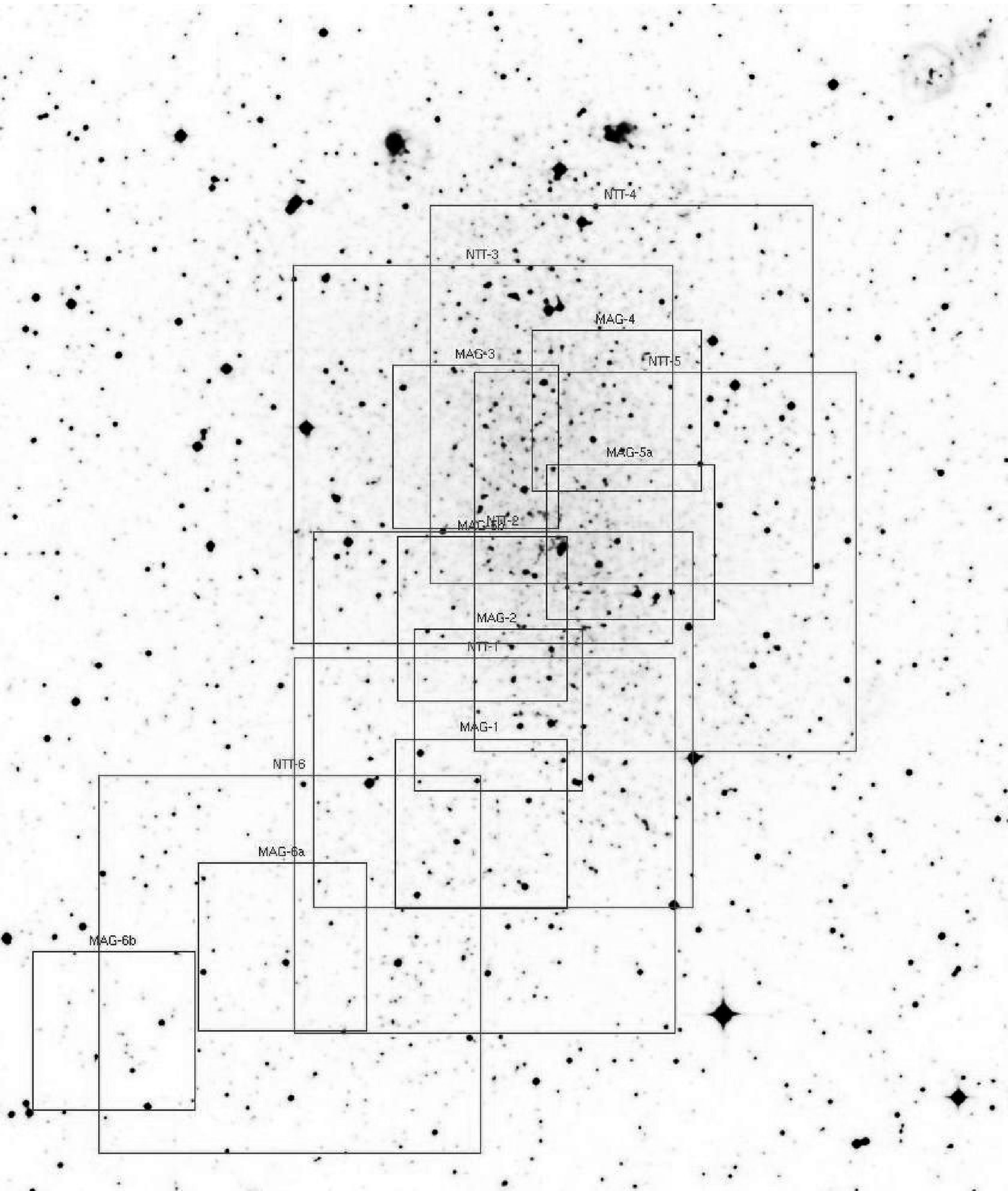} 
\caption{The location of the observed fields in NGC 6822 on the DSS
blue plate. We observed six NTT/SOFI fields, and eight Magellan/PANIC fields (see text).
The fields cover most of the spatial extent of the galaxy and contain 17 Cepheid
variables with periods in excess of 10 days.}
\end{figure}

\begin{figure}[p]
\vspace*{18cm}
\includegraphics{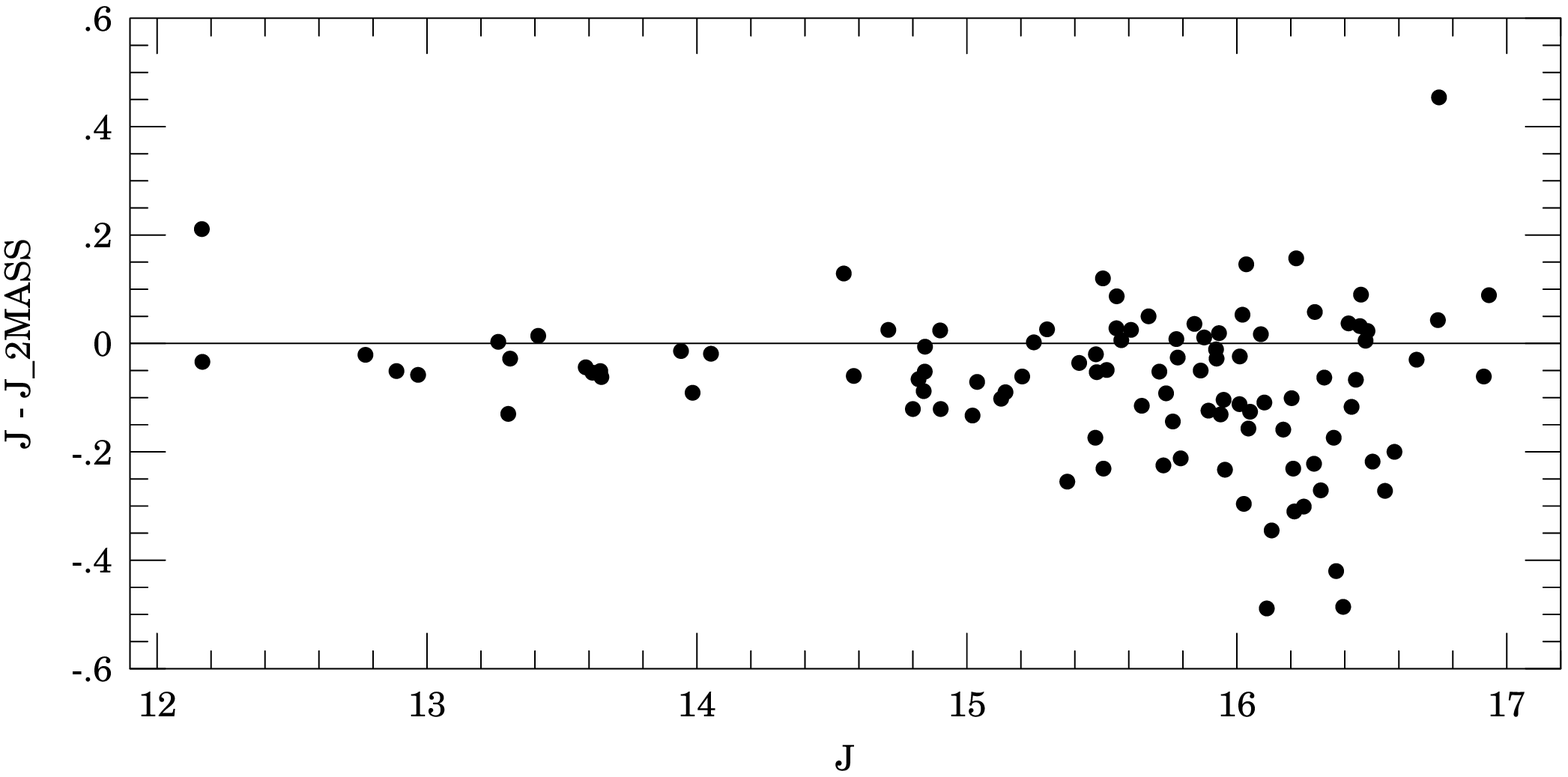}
\includegraphics{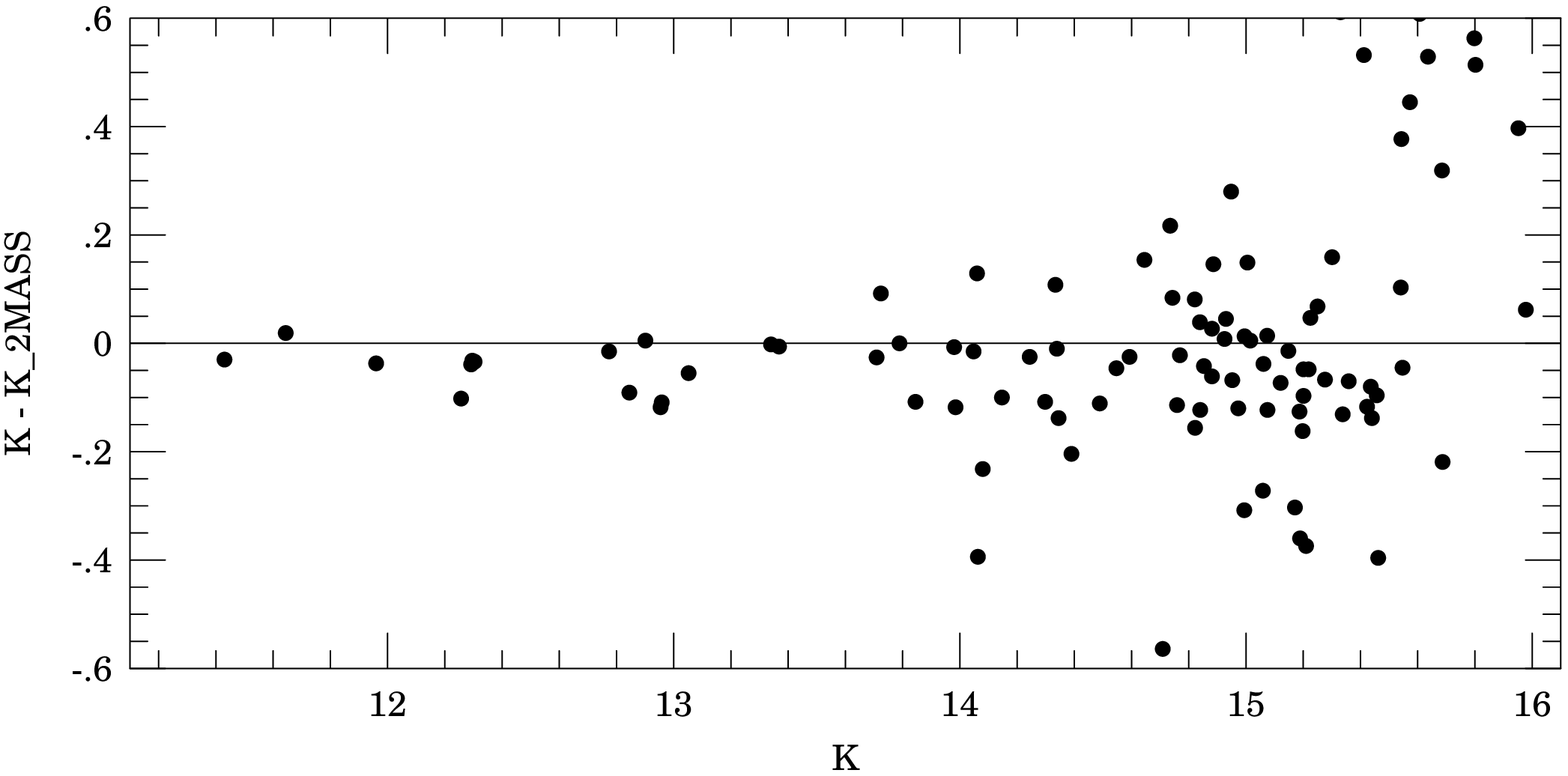}
\vspace{-3cm}
\caption{Comparison of our NGC 6822 near-infared photometry with 2MASS data for common stars.
The brightest common stars with reasonably high S/N in the 2MASS photometry suggest
a slight systematic offset of -0.03 mag in both bands in the sense that
our J and K magnitudes are on average brighter by this amount. The 
deviation is likely to be not significant.}
\end{figure}

\begin{figure}[p]
\includegraphics{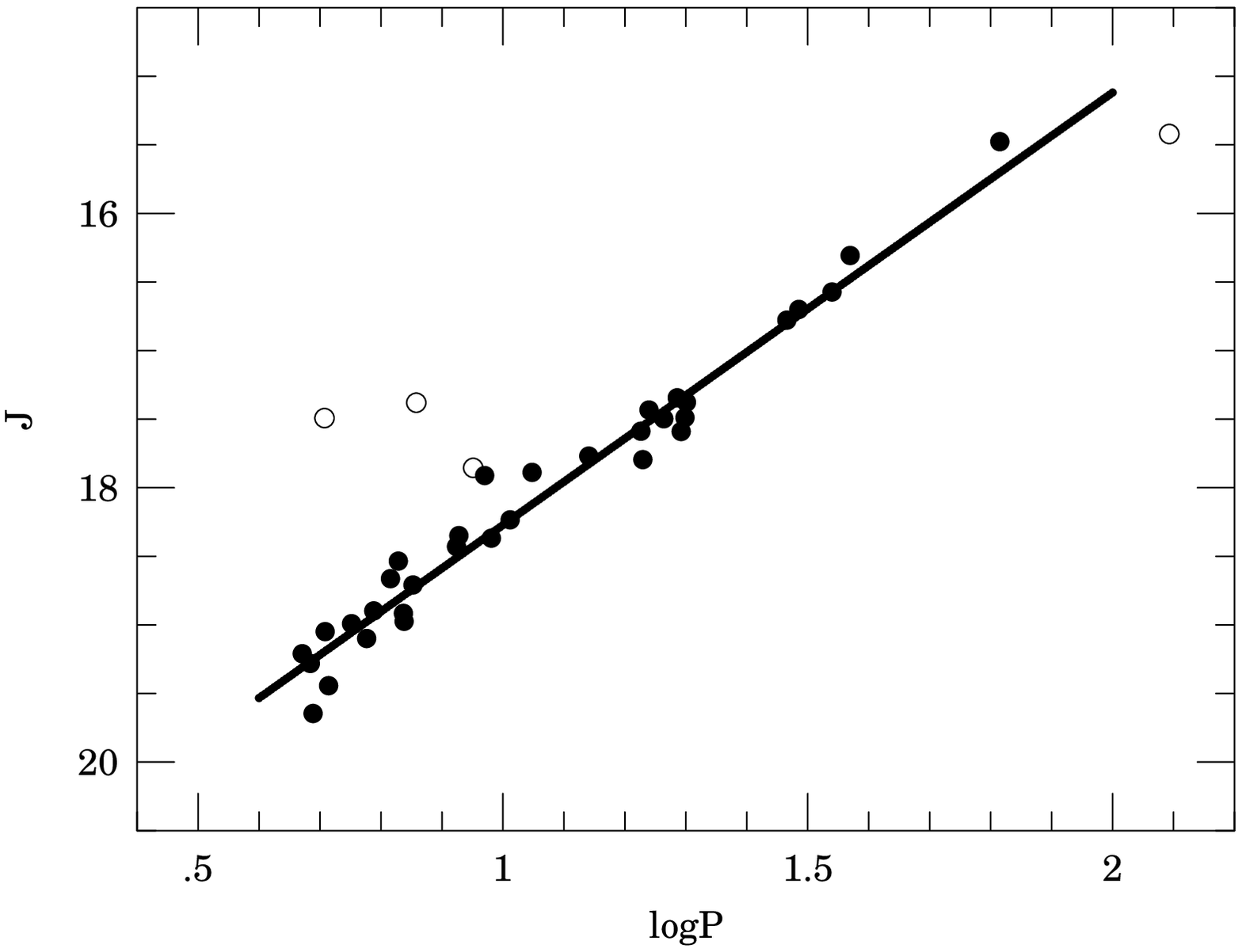}
\includegraphics{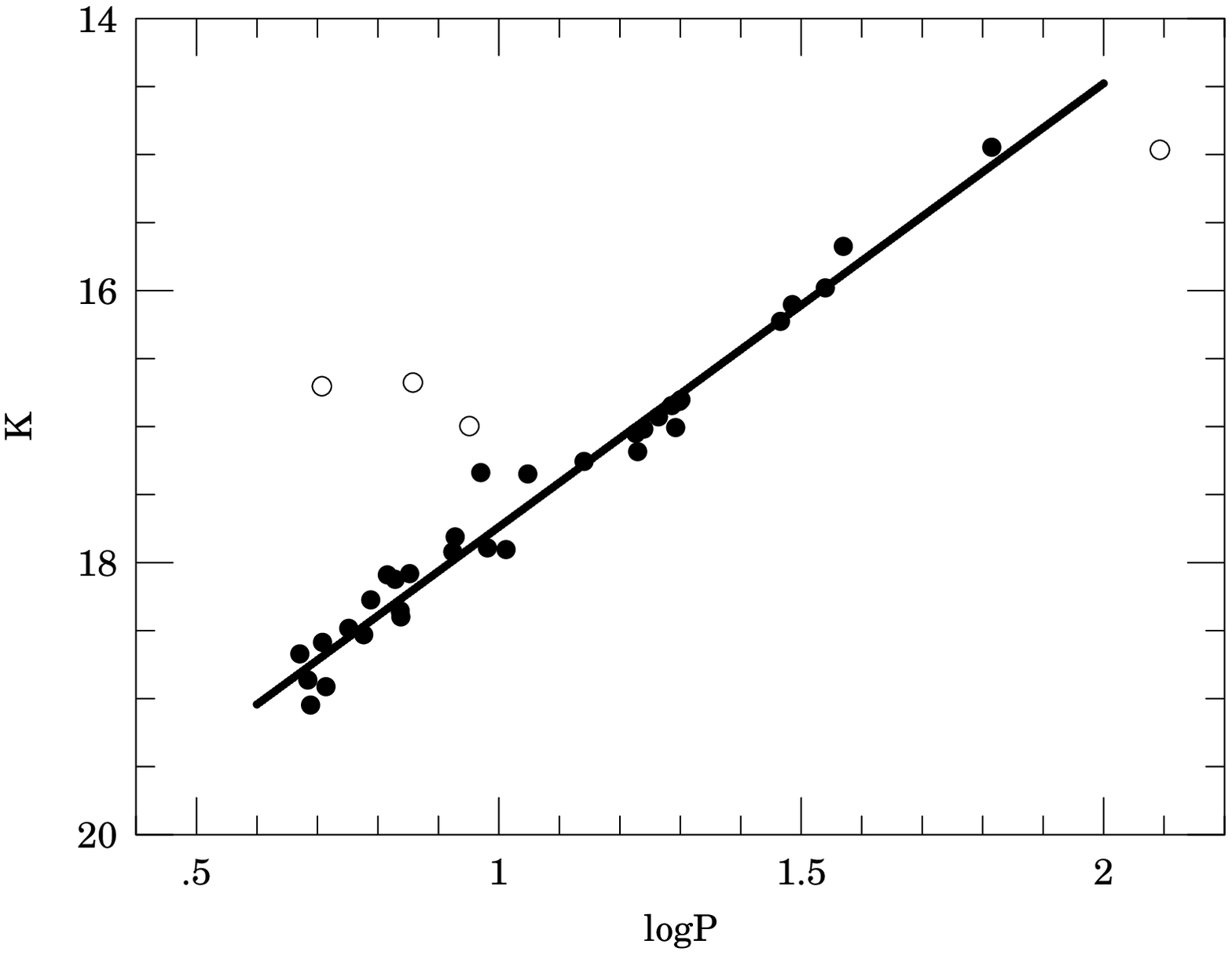}
\vspace{22cm} 
\caption{Cepheid period-luminosity relations for NGC 6822 in the J (upper panel), 
and K (lower panel) bands. We have excluded the 4 objects indicated by open circles
in these figures in the distance solutions, for the reasons given in the
text. We have also omitted all variables with periods shorter than our
adopted cutoff period of 4.8 days (see text).
}
\end{figure}

\begin{figure}[p]
\includegraphics{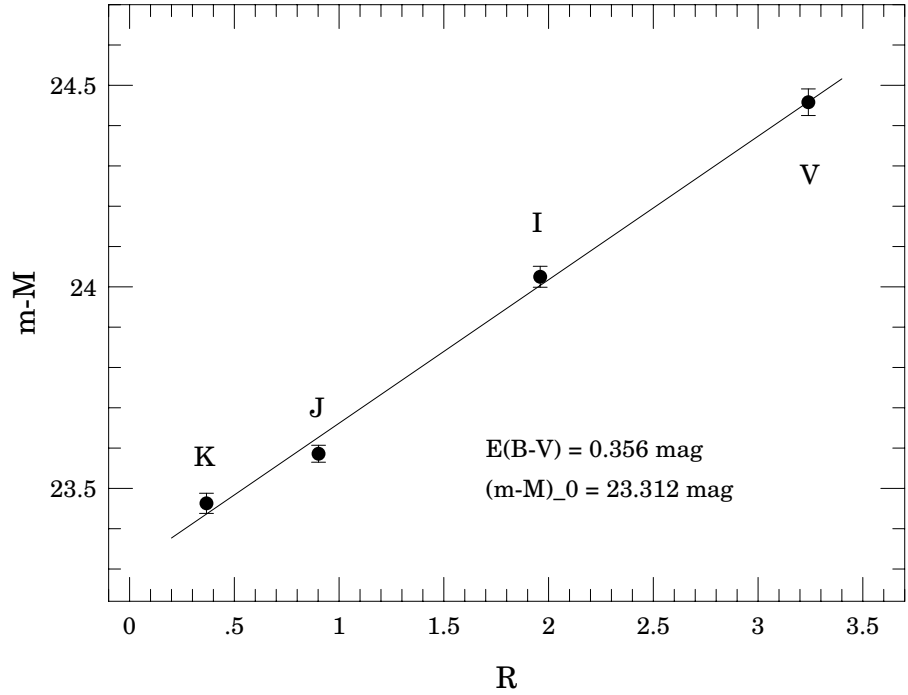}
\vspace{10cm}
\caption{Apparent distance moduli to NGC 6822 as derived in different photometric bands,
plotted against the ratio of total to selective extinction as adopted from
the Schlegel et al. reddening law. The intersection and
slope of the best-fitting line give the true distance modulus and
reddening, respectively. The data in this diagram suggest that the galactic reddening law
is a very good approximation for NGC 6822 as well.}
\end{figure}

\clearpage
\begin{deluxetable}{c c c c c c c}
%\rotate
\tablewidth{0pc}
\tablecaption{Journal of the Individual J and K band Observations of NGC 6822
Cepheids}
\tablehead{ \colhead{ID} & \colhead{J HJD} & \colhead{J}  & \colhead{$\sigma$} 
& \colhead{K HJD} & \colhead{K} & \colhead{$\sigma$} }
\startdata
cep001 & 3215.771560 &  15.515 &    0.026 & 3215.716360 &  14.990 &   0.010\\ 
cep001 & 3226.567980 &  15.517 &    0.009 & 3226.502680 &  14.932 &   0.021\\ 
cep001 & 3250.648500 &  15.543 &    0.008 & 3250.664120 &  15.029 &   0.009\\ 
cep002 & 3215.619160 &  15.555 &    0.005 & 3215.564180 &  14.948 &   0.005\\ 
cep002 & 3215.692830 &  15.544 &    0.004 & 3215.636120 &  14.947 &   0.005\\ 
cep002 & 3250.545020 &  15.347 &    0.005 & 3250.571990 &  14.905 &   0.009\\ 
cep003 & 3215.771560 &  16.396 &    0.042 & 3215.716360 &  15.892 &   0.021\\ 
cep003 & 3250.724570 &  16.526 &    0.009 & 3250.694500 &  15.809 &   0.018\\ 
cep004 & 3215.619160 &  16.597 &    0.012 & 3215.564180 &  15.941 &   0.008\\ 
cep004 & 3215.692830 &  16.639 &    0.010 & 3215.636120 &  15.841 &   0.023\\ 
cep004 & 3250.633530 &  16.581 &    0.006 & 3250.602610 &  15.917 &   0.012\\ 
cep007 & 3215.771560 &  16.787 &    0.013 & 3215.716360 &  16.120 &   0.016\\ 
cep007 & 3226.567980 &  16.750 &    0.014 & 3226.502680 &  16.327 &   0.021\\ 
cep007 & 3226.664610 &  16.673 &    0.014 & 3226.603750 &  16.215 &   0.028\\ 
cep007 & 3250.648500 &  17.014 &    0.009 & 3250.664120 &  16.309 &   0.015\\ 
cep008 & 3215.771560 &  16.577 &    0.015 & 3215.716360 &  16.073 &   0.025\\ 
cep008 & 3226.567980 &  16.717 &    0.015 & 3226.502680 &  16.135 &   0.015\\ 
cep008 & 3226.664610 &  16.655 &    0.014 & 3226.603750 &  16.178 &   0.018\\ 
cep008 & 3250.724570 &  16.805 &    0.007 & 3250.694500 &  16.142 &   0.014\\ 
cep010 & 3215.619160 &  17.364 &    0.009 & 3215.564180 &  16.726 &   0.009\\ 
cep010 & 3215.692830 &  17.379 &    0.010 & 3215.636120 &  16.711 &   0.011\\ 
cep010 & 3250.545020 &  17.238 &    0.009 & 3250.571990 &  16.650 &   0.016\\ 
cep011 & 3215.771560 &  17.410 &    0.014 & 3215.716360 &  16.813 &   0.018\\ 
cep011 & 3250.724570 &  17.630 &    0.015 & 3250.694500 &  16.929 &   0.025\\ 
cep012 & 3215.619160 &  17.717 &    0.012 & 3215.564180 &  17.075 &   0.014\\ 
cep012 & 3215.692830 &  17.741 &    0.018 & 3215.636120 &  17.097 &   0.016\\ 
cep013 & 3251.560460 &  17.212 &    0.031 & 3251.528590 &  16.790 &   0.031\\ 
cep014 & 3215.619160 &  17.668 &    0.017 & 3215.564180 &  17.020 &   0.019\\ 
cep014 & 3215.692830 &  17.715 &    0.014 & 3215.636120 &  17.067 &   0.017\\ 
cep015 & 3215.619160 &  17.631 &    0.023 & 3215.564180 &  17.150 &   0.027\\ 
cep015 & 3215.692830 &  17.631 &    0.018 & 3215.636120 &  17.138 &   0.015\\ 
cep015 & 3226.664610 &  17.316 &    0.015 & 3226.603750 &  16.895 &   0.023\\ 
cep016 & 3215.619160 &  17.658 &    0.015 & 3215.564180 &  17.074 &   0.012\\ 
cep017 & 3215.771560 &  17.421 &    0.021 & 3215.716360 &  16.908 &   0.030\\ 
cep017 & 3251.560460 &  17.621 &    0.016 & 3251.528590 &  17.005 &   0.024\\ 
cep018 & 3215.619160 &  17.925 &    0.016 & 3215.564180 &  17.371 &   0.021\\ 
cep018 & 3215.692830 &  18.008 &    0.016 & 3215.636120 &  17.431 &   0.023\\ 
cep018 & 3250.633530 &  17.631 &    0.016 & 3250.602610 &  17.103 &   0.020\\ 
cep018 & 3299.592960 &  17.801 &    0.026 & 3299.551310 &  17.445 &   0.018\\ 
cep019 & 3215.692830 &  17.913 &    0.016 & 3215.636120 &  17.276 &   0.021\\ 
cep019 & 3215.771560 &  18.151 &    0.020 & 3215.716360 &  17.598 &   0.026\\ 
cep019 & 3226.567980 &  17.874 &    0.031 & 3226.502680 &  17.454 &   0.042\\ 
cep019 & 3226.664610 &  17.977 &    0.020 & 3226.603750 &  17.329 &   0.025\\ 
cep019 & 3251.560460 &  17.846 &    0.021 & 3251.528590 &  17.271 &   0.030\\ 
cep022 & 3215.692830 &  18.268 &    0.037 & 3215.636120 &  17.862 &   0.052\\ 
cep023 & 3215.619160 &  18.266 &    0.018 & 3215.564180 &  17.822 &   0.025\\ 
cep024 & 3226.567980 &  17.594 &    0.021 & 3226.502680 &  17.027 &   0.032\\ 
cep024 & 3226.664610 &  17.678 &    0.020 & 3226.603750 &  17.064 &   0.042\\ 
cep024 & 3251.560460 &  18.128 &    0.018 & 3251.528590 &  17.738 &   0.037\\ 
cep025 & 3215.771560 &  17.915 &    0.029 & 3215.716360 &  17.047 &   0.016\\ 
cep025 & 3250.648500 &  17.877 &    0.015 & 3250.664120 &  16.933 &   0.024\\ 
cep026 & 3215.619160 &  18.304 &    0.016 & 3215.564180 &  17.765 &   0.022\\ 
cep026 & 3215.692830 &  18.239 &    0.016 & 3215.636120 &  17.797 &   0.026\\ 
cep026 & 3250.545020 &  18.188 &    0.020 & 3250.571990 &  17.662 &   0.044\\ 
cep027 & 3215.619160 &  18.353 &    0.027 & 3215.564180 &  17.847 &   0.030\\ 
cep027 & 3215.692830 &  18.336 &    0.025 & 3215.636120 &  17.859 &   0.030\\ 
cep028 & 3215.619160 &  17.353 &    0.012 & 3215.564180 &  16.591 &   0.013\\ 
cep028 & 3215.692830 &  17.351 &    0.011 & 3215.636120 &  16.675 &   0.016\\ 
cep029 & 3215.619160 &  18.710 &    0.025 & 3215.564180 &  18.010 &   0.031\\ 
cep029 & 3215.692830 &  18.679 &    0.025 & 3215.636120 &  18.000 &   0.034\\ 
cep029 & 3250.633530 &  18.652 &    0.023 & 3250.602610 &  18.284 &   0.048\\ 
cep029 & 3250.545020 &  18.744 &    0.027 & 3250.571990 &  18.237 &   0.044\\ 
cep030 & 3215.771560 &  19.090 &    0.051 & 3215.716360 &  18.519 &   0.047\\ 
cep030 & 3250.724570 &  19.041 &    0.065 & 3250.694500 &  18.297 &   0.056\\ 
cep031 & 3215.771560 &  18.872 &    0.030 & 3215.716360 &  18.256 &   0.041\\ 
cep031 & 3250.724570 &  19.042 &    0.028 & 3250.694500 &  18.386 &   0.044\\ 
cep033 & 3215.771560 &  18.367 &    0.034 & 3215.716360 &  17.838 &   0.051\\ 
cep033 & 3299.592960 &  18.761 &    0.027 & 3299.551310 &  18.392 &   0.036\\ 
cep034 & 3215.692830 &  18.653 &    0.029 & 3215.636120 &  18.098 &   0.036\\ 
cep034 & 3215.771560 &  18.496 &    0.023 & 3215.716360 &  17.976 &   0.031\\ 
cep037 & 3250.724570 &  18.985 &    0.036 & 3250.694500 &  18.289 &   0.047\\ 
cep041 & 3215.692830 &  19.329 &    0.047 & 3215.636120 &  18.599 &   0.046\\ 
cep041 & 3215.771560 &  19.340 &    0.050 & 3215.716360 &  18.622 &   0.047\\ 
cep041 & 3299.592960 &  19.014 &    0.043 & 3299.551310 &  18.513 &   0.044\\ 
cep043 & 3250.648500 &  18.858 &    0.021 & 3250.664120 &  18.410 &   0.042\\ 
cep048 & 3250.724570 &  19.459 &    0.035 & 3250.694500 &  18.984 &   0.069\\ 
cep051 & 3215.619160 &  19.032 &    0.033 & 3215.564180 &  18.540 &   0.050\\ 
cep051 & 3215.692830 &  18.984 &    0.033 & 3215.636120 &  18.566 &   0.041\\ 
cep051 & 3250.633530 &  18.875 &    0.019 & 3250.602610 &  18.549 &   0.044\\ 
cep052 & 3215.771560 &  17.435 &    0.024 & 3215.716360 &  16.576 &   0.014\\ 
cep056 & 3215.771560 &  19.536 &    0.068 & 3215.716360 &  19.015 &   0.088\\ 
cep057 & 3215.619160 &  19.159 &    0.031 & 3215.564180 &  18.855 &   0.053\\ 
cep058 & 3215.692830 &  19.181 &    0.062 & 3215.636120 &  18.596 &   0.055\\ 
cep058 & 3215.771560 &  19.234 &    0.060 & 3215.716360 &  18.620 &   0.054\\ 
cep058 & 3251.560460 &  19.248 &    0.030 & 3251.528590 &  18.803 &   0.056\\ 
cep061 & 3215.771560 &  18.687 &    0.037 & 3215.716360 &  17.949 &   0.045\\ 
cep063 & 3215.771560 &  18.605 &    0.031 & 3215.716360 &  18.061 &   0.044\\ 
cep063 & 3250.724570 &  18.823 &    0.026 & 3250.694500 &  18.178 &   0.043\\ 
cep064 & 3215.619160 &  18.852 &    0.028 & 3215.564180 &  18.390 &   0.038\\ 
cep064 & 3215.692830 &  19.016 &    0.034 & 3215.636120 &  18.356 &   0.033\\ 
cep067 & 3215.619160 &  19.490 &    0.038 & 3215.564180 &  19.024 &   0.059\\ 
cep068 & 3215.619160 &  18.627 &    0.035 & 3215.564180 &  17.918 &   0.085\\ 
cep068 & 3215.692830 &  18.668 &    0.029 & 3215.636120 &  17.955 &   0.039\\ 
cep068 & 3226.664610 &  18.460 &    0.041 & 3226.603750 &  17.929 &   0.053\\ 
cep069 & 3215.692830 &  19.176 &    0.043 & 3215.636120 &  19.171 &   0.070\\ 
cep070 & 3215.692830 &  19.551 &    0.070 & 3215.636120 &  18.842 &   0.096\\ 
cep073 & 3215.619160 &  20.156 &    0.129 & 3215.564180 &  19.485 &   0.149\\ 
cep075 & 3226.567980 &  17.322 &    0.019 & 3226.502680 &  15.978 &   0.013\\ 
cep076 & 3215.692830 &  18.350 &    0.035 & 3215.636120 &  17.897 &   0.070\\ 
cep076 & 3299.592960 &  18.586 &    0.062 & 3299.551310 &  18.333 &   0.072\\ 
cep077 & 3215.619160 &  18.972 &    0.053 & 3215.564180 &  18.300 &   0.067\\ 
cep077 & 3215.692830 &  18.965 &    0.041 & 3215.636120 &  18.420 &   0.068\\ 
cep077 & 3250.633530 &  18.958 &    0.031 & 3250.602610 &  18.589 &   0.047\\ 
cep078 & 3250.633530 &  19.084 &    0.030 & 3250.602610 &  18.727 &   0.062\\ 
cep078 & 3299.592960 &  19.090 &    0.032 & 3299.551310 &  18.870 &   0.043\\ 
cep083 & 3215.619160 &  19.311 &    0.046 & 3215.564180 &  18.454 &   0.058\\ 
cep097 & 3215.619160 &  19.412 &    0.041 & 3215.564180 &  19.117 &   0.063\\ 
cep097 & 3215.692830 &  19.497 &    0.042 & 3215.636120 &  18.972 &   0.054\\ 
cep097 & 3250.633530 &  19.361 &    0.023 & 3250.602610 &  18.889 &   0.062\\ 
cep097 & 3250.545020 &  19.319 &    0.025 & 3250.571990 &  19.009 &   0.067\\ 
cep099 & 3250.545020 &  19.847 &    0.038 & 3250.571990 &  19.679 &   0.108\\ 
cep101 & 3215.619160 &  18.536 &    0.028 & 3215.564180 &  17.608 &   0.021\\ 
cep101 & 3215.692830 &  18.505 &    0.028 & 3215.636120 &  17.576 &   0.020\\ 
cep103 & 3215.692830 &  19.401 &    0.058 & 3215.636120 &  18.401 &   0.054\\ 
cep113 & 3215.619160 &  19.074 &    0.032 & 3215.564180 &  18.595 &   0.042\\ 
cep116 & 3215.619160 &  18.980 &    0.039 & 3215.564180 &  18.220 &   0.040\\ 
cep116 & 3215.692830 &  18.982 &    0.037 & 3215.636120 &  18.193 &   0.037\\ 
\enddata
\end{deluxetable}

\clearpage

\begin{deluxetable}{c c c c c c}
\tablecaption{Intensity mean J and K magnitudes for 56 Cepheid variables in NGC 6822}
%\tablewidth{0pt}
\tablehead{
\colhead{ID} & \colhead{log P} &
\colhead{$<J>$} & \colhead{$\sigma_{\rm J}$} & \colhead{$<K>$} &
\colhead{$\sigma_{\rm K}$}\\ 
& \colhead{days} & \colhead{mag} & \colhead{mag} & \colhead{mag} &
\colhead{mag}
}
\startdata
cep001 &    2.093 &  15.445 &    0.030 &  14.964 &   0.029\\ 
cep002 &    1.815 &  15.475 &    0.025 &  14.946 &   0.026\\ 
cep003 &    1.570 &  16.277 &    0.039 &  15.678 &   0.032\\ 
cep004 &    1.540 &  16.579 &    0.027 &  15.963 &   0.030\\ 
cep007 &    1.485 &  16.695 &    0.028 &  16.099 &   0.032\\ 
cep008 &    1.466 &  16.740 &    0.028 &  16.219 &   0.031\\ 
cep010 &    1.301 &  17.376 &    0.027 &  16.802 &   0.028\\ 
cep011 &    1.299 &  17.489 &    0.029 &  16.812 &   0.033\\ 
cep012 &    1.292 &  17.592 &    0.029 &  17.008 &   0.029\\ 
cep013 &    1.286 &  17.344 &    0.040 &  16.845 &   0.040\\ 
cep014 &    1.264 &  17.495 &    0.029 &  16.927 &   0.031\\ 
cep015 &    1.240 &  17.435 &    0.031 &  17.020 &   0.033\\ 
cep016 &    1.229 &  17.795 &    0.029 &  17.184 &   0.028\\ 
cep017 &    1.226 &  17.579 &    0.031 &  17.047 &   0.037\\ 
cep018 &    1.141 &  17.768 &    0.031 &  17.254 &   0.032\\ 
cep019 &    1.048 &  17.884 &    0.033 &  17.355 &   0.039\\ 
cep022 &    1.012 &  18.234 &    0.045 &  17.901 &   0.053\\ 
cep023 &    0.981 &  18.368 &    0.031 &  17.892 &   0.035\\ 
cep024 &    0.972 &  17.896 &    0.032 &  17.342 &   0.045\\ 
cep025 &    0.951 &  17.854 &    0.034 &  16.987 &   0.032\\ 
cep026 &    0.928 &  18.345 &    0.030 &  17.802 &   0.041\\ 
cep027 &    0.924 &  18.429 &    0.036 &  17.921 &   0.039\\ 
cep028 &    0.858 &  17.380 &    0.028 &  16.680 &   0.029\\ 
cep029 &    0.852 &  18.711 &    0.035 &  18.099 &   0.047\\ 
cep030 &    0.838 &  18.968 &    0.064 &  18.387 &   0.057\\ 
cep031 &    0.837 &  18.915 &    0.038 &  18.355 &   0.049\\ 
cep033 &    0.828 &  18.521 &    0.040 &  18.090 &   0.051\\ 
cep034 &    0.816 &  18.672 &    0.036 &  18.094 &   0.042\\ 
cep037 &    0.788 &  18.898 &    0.044 &  18.274 &   0.053\\ 
cep041 &    0.776 &  19.109 &    0.053 &  18.532 &   0.052\\ 
cep043 &    0.752 &  18.991 &    0.033 &  18.483 &   0.049\\ 
cep048 &    0.714 &  19.443 &    0.043 &  18.912 &   0.073\\ 
cep051 &    0.708 &  19.071 &    0.038 &  18.585 &   0.052\\ 
cep052 &    0.707 &  17.492 &    0.035 &  16.703 &   0.029\\ 
cep056 &    0.689 &  19.646 &    0.072 &  19.047 &   0.091\\ 
cep057 &    0.684 &  19.281 &    0.040 &  18.863 &   0.059\\ 
cep058 &    0.671 &  19.233 &    0.058 &  18.672 &   0.060\\ 
cep061 &    0.658 &  18.821 &    0.045 &  18.028 &   0.051\\ 
cep063 &    0.648 &  18.698 &    0.038 &  18.140 &   0.050\\ 
cep064 &    0.639 &  19.060 &    0.040 &  18.445 &   0.043\\ 
cep067 &    0.627 &  19.499 &    0.045 &  19.079 &   0.064\\ 
cep068 &    0.626 &  18.588 &    0.043 &  17.967 &   0.067\\ 
cep069 &    0.625 &  19.355 &    0.050 &  19.294 &   0.074\\ 
cep070 &    0.614 &  19.435 &    0.074 &  18.665 &   0.099\\ 
cep073 &    0.601 &  20.248 &    0.131 &  19.435 &   0.151\\ 
cep075 &    0.591 &  17.233 &    0.031 &  15.986 &   0.028\\ 
cep076 &    0.578 &  18.515 &    0.056 &  18.155 &   0.075\\ 
cep077 &    0.577 &  19.062 &    0.049 &  18.505 &   0.066\\ 
cep078 &    0.570 &  19.171 &    0.040 &  18.782 &   0.059\\ 
cep083 &    0.554 &  19.418 &    0.052 &  18.564 &   0.063\\ 
cep097 &    0.455 &  19.368 &    0.042 &  18.952 &   0.067\\ 
cep099 &    0.426 &  20.010 &    0.045 &  19.709 &   0.111\\ 
cep101 &    0.414 &  18.535 &    0.038 &  17.639 &   0.032\\ 
cep103 &    0.405 &  19.261 &    0.063 &  18.349 &   0.060\\ 
cep113 &    0.316 &  19.119 &    0.041 &  18.605 &   0.049\\ 
cep116 &    0.233 &  19.047 &    0.045 &  18.239 &   0.046\\ 
\enddata
\end{deluxetable}

\begin{deluxetable}{cccccc}
\tablewidth{0pc}
\tablecaption{Reddened and Absorption-Corrected Distance Moduli for NGC
6822 in Optical and Near-Infrared Bands}
\tablehead{ \colhead{Band} & $V$ & $I$ & $J$ & $K$ & $E(B-V)$ }
\startdata
 $m-M$                &   24.458 &  24.025 &  23.586 &  23.463 &   --  \nl
 ${\rm R}_{\lambda}$  &   3.24   &  1.96   &  0.902  &  0.367  &   --  \nl
$(m-M)_{0}$           &   23.304 &  23.327 &  23.264 &  23.332 &  0.356 \nl
\enddata
\end{deluxetable}

\end{document}